\documentclass[aps,twocolumn,showpacs,superscriptaddress]{revtex4}
\usepackage{graphicx}

\begin{document}

\title{Survival of quantum effects for observables after decoherence} 
\author{G.P. Berman}
\affiliation{Theoretical Division and CNLS,MS B213, 
Los Alamos National Laboratory, Los Alamos, NM 87545} 

\author{A.R. Bishop}
\affiliation{Theoretical Division, MS B210, Los Alamos National Laboratory, 
Los Alamos, NM 87545} 

\author{F. Borgonovi}
\affiliation{Dipartimento di Matematica e Fisica, Universit\`a Cattolica, 
via Musei 41, 25121 Brescia, Italy}
\affiliation{INFM, Unit\`a di Brescia and INFN, Sezione di Pavia, Italy}

\author{D.A.R. Dalvit}
\affiliation{Theoretical Division, MS B213, Los Alamos National Laboratory, 
Los Alamos, NM 87545}

\date{\today} 

\begin{abstract} 

{When a quantum nonlinear system is linearly coupled to an infinite bath
of harmonic oscillators, quantum coherence of the system is lost on a 
decoherence time-scale $\tau_D$. Nevertheless, quantum effects for observables may still survive
environment-induced decoherence, and be observed for times much larger than the
decoherence time-scale. In particular, we show that the Ehrenfest time, 
which characterizes a  departure of quantum dynamics for observables 
from the corresponding classical dynamics, can be observed for a 
quasi-classical nonlinear oscillator for times $\tau \gg\tau_D$. 
We discuss this observation in relation to recent 
experiments on quantum nonlinear systems in the quasi-classical region of 
parameters.}

\end{abstract} 

\maketitle 

In the last few decades there has been extensive theoretical, and more recently 
experimental, research on the quantum-classical transition. It has been noted that 
every physical system is, in fact, an open quantum system interacting with its 
environment. Consequently, the evolution of the reduced density matrix of the 
system (obtained after tracing over the environmental variables) evolves in such a way 
that quantum coherent effects are quickly suppressed. This process of 
environment-induced decoherence has been considered to be an essential ingredient 
of the quantum-classical transition \cite{decoherence}. On the other hand, despite 
the huge number of papers on this subject, only few of them deal with 
quantum nonlinear systems (e.g. \cite{haake,milburn1,milburn2,milburn3,zurekpaz}).

We consider in this paper the dynamics of a quantum nonlinear oscillator (QNO)
\begin{equation}
\hat H = \hbar \omega \hat a^{\dagger} 
\hat a + \mu \hbar^2 (\hat a^{\dagger} \hat a)^2 ,
\label{Hnonlinear}
\end{equation}
interacting with a bath of linear oscillators which are initially in thermal equilibrium. 
Here $\hat a~ ({\hat a}^{\dagger})$ are annihilation (creation) bosonic operators, 
$\omega$ is the linear frequency and $\mu$ is the parameter of nonlinearity.
The QNO is initially prepared in a coherent state in the quasi-classical region 
of parameters. 
In the classical limit ($\hat a\rightarrow\alpha$, $\hat a^\dagger\rightarrow\alpha^*$, 
$\hbar\rightarrow 0$, $|\alpha|\rightarrow\infty$, $\hbar|\alpha|^2=J$ - an action of the 
classical linear oscillator) the Hamiltonian (\ref{Hnonlinear}) becomes 
$H_{\rm cl}=\omega J+\mu J^2$. In what follows we use the following dimensionless 
notation: $\tau=\omega t$, $\bar\mu=\hbar\mu/\omega$, $\mu_{\rm cl}=\mu J/\omega$ and 
$\varepsilon=\hbar/J$. Thus, the quantum parameter $\bar\mu$ can be presented as a 
product of two parameters, quantum and classical, $\bar\mu=\varepsilon\mu_{\rm cl}$. 
The parameter $\mu_{\rm cl}$ characterizes the nonlinearity in the classical system, 
and can be written as $\mu_{\rm cl}=(J/2 \omega) (d\omega_{\rm cl}/dJ)$, where 
$\omega_{\rm cl}=dH_{\rm cl}/dJ=\omega+2\mu J$ is the classical frequency of 
nonlinear oscillations. The limit $\mu_{\rm cl}\ll 1$ corresponds to  weak nonlinearity,
while $\mu_{\rm cl}\ge 1$ corresponds to strong nonlinearity. 
The parameter $\varepsilon$ is a quasi-classical parameter. Namely, 
$\varepsilon\sim 1$ corresponds to the ``pure'' quantum system, and 
$\varepsilon\ll 1$ corresponds to the quasi-classical limit, which is the 
subject of this paper.  

We study the following problem: What are the parameter conditions 
for observation of quantum effects on expectation values (observables) 
in the QNO dynamics. 
We describe the dynamics for the QNO for observables
taking into account five characteristic time-scales which naturally appear in 
this system. Three of them characterize the time-scales of the QNO evolving
under the Hamiltonian dynamics: 
(i) $\tau_{\rm cl}=2 \pi / \omega_{\rm cl}$ -- the period of nonlinear classical oscillations; 
(ii) $\tau_E$ -- the so-called Ehrenfest time, which indicates the 
characteristic time-scale at which quantum dynamics for observables starts to 
depart from the corresponding classical dynamics; 
(iii) $\tau_R$ -- a quantum recurrence time, which describes the time-scale for
quantum recurrences of  observables under the Hamiltonian evolution. 
There are also two characteristic time-scales related to the interaction of the 
QNO with the thermal bath: 
(iv) $\tau_D$ -- a decoherence time, which characterizes the decay of the 
non-diagonal matrix elements of the reduced density matrix in the  
eigenbasis of the non-interacting Hamiltonian; and 
(v) $\tau_\gamma$ -- a time-scale of relaxation of quantum observables due to 
the interaction with the thermal bath. 
We demonstrate that even if the decoherence time is much smaller than the 
Ehrenfest time, $\tau_D\ll\tau_E$, one can still observe quantum effects for observables. Actually, 
the important condition for observation of quantum effects related to the Ehrenfest 
time-scale is $\tau_E<\tau_\gamma$, which may be realized in modern 
experiments in the quasi-classical region of parameters ($\varepsilon\ll 1$). 
This means that  generally the 
environment-induced decoherence is insufficient for recovering
the quantum-classical correspondence for observables in quantum 
nonlinear systems. This is an important observation for at least two reasons: 
(a) It means that pure quantum effects can be observed for times much longer than 
$\tau_D$ and (b) Pure quantum dynamical effects can be important in experiments 
even in the quasi-classical region of parameters. Finally, the classical limit 
appears in our system under very natural conditions:   
$\tau_D \ll \tau_{\rm cl} \ll \tau_{\gamma} \ll \tau_E \ll \tau_R$. 

One type of systems that could be considered for observing quantum effects in the 
quasi-classical region of parameters are Bose-Einstein condensates (BECs).
These are particularly suited to analyzing the interplay between nonlinear 
dynamics and environmental interactions in the realm of quantum mesoscopic systems. 
This is because they are macroscopic matter waves, often described theoretically by 
the Gross-Pitaevskii (GP) equation, which formally is a classical nonlinear field-theory. 
Going beyond GP allows one to understand the role of quantum effects in the 
quasi-classical region of parameters.
Another important feature of BECs is that they are 
experimentally easily accessible and controllable by means of trapping potentials
and tunable interactions. BECs have already been used to demonstrate nonlinear
dynamics of collapses
and revivals of a coherent matter wave in the pure quantum regime 
\cite{greiner}. Here, the condensate was initially trapped 
in the lowest energy band of a three-dimensional optical lattice. 
By adiabatically raising the heights of the barriers it was 
possible to suppress tunneling between sites and at the same time 
maintain the system in the superfluid regime, so that 
within each lattice site independent coherent states were 
engineered with an average number of atoms of order one.
Every BEC in each lattice site is well described in the single-mode 
approximation by the QNO Hamiltonian given in Eq. (\ref{Hnonlinear}),
with $\omega$ being the trapping frequency in each lattice site and 
$\mu$ being proportional to the s-wave two-body scattering length.
Other systems that could be used for observation of quantum nonlinear effects 
in the quasi-classical region of parameters are micro- \cite{chan}
and nanomechanical \cite{roukes} resonators, high-frequency cantilevers \cite{rugar},
nonlinear optical systems and superconductive junctions.

The QNO is one of the simplest quantum nonlinear systems for 
which the breakdown 
of the quantum-classical correspondence can be exactly calculated. The quantum and 
classical dynamics of an initial coherent wave packet evolving 
under Eq. (\ref{Hnonlinear}) were computed by Berman {\it et al} \cite{genna1} 
and  by Milburn \cite{milburn1}. The characteristic time-scale for departure 
of the quantum dynamics from the corresponding classical one, the so-called Ehrenfest
time, was introduced for this system in \cite{genna1}. In \cite{milburn1} a similar 
problem was 
studied using the $Q$ quasi-probability distribution.
It was shown that the presence of non-positive definite 
second-order terms in the quantum evolution equation for $Q$, not present in the 
evolution of the classical probability density, 
are responsible for quantum recurrences and prevent the 
appearance of fine-scale-structure ``whorls'' predicted in the classical 
description. In \cite{milburn2,milburn3} the interaction of the QNO with the environment 
(modeled by a thermal bath of harmonic oscillators) was studied in the limit of small nonlinearity,
and it was argued that such interaction 
was effective in destroying quantum interference effects and restoring the 
classical phase-space structure. However, as we already stated, environment-induced
decoherence is in fact ineffective in recovering the quantum-classical correspondence
for this nonlinear system.

In the following we compute quantum observables in 
the coherent state basis.  
For an arbitrary operator function 
$\hat f = \hat f(\hat a^{\dagger}, \hat a)$, the time-dependent 
expectation value
of such a function (observable), 
$f(\alpha^{\star},\alpha,t)= 
\langle \alpha|e^{i\hat Ht/\hbar}\hat f e^{-i\hat H t/\hbar} |\alpha \rangle$, 
for an initial coherent state, $|\alpha \rangle$,  satisfies a 
partial differential equation of the form \cite{gennabook}:
$\partial f / \partial \tau = \hat K  f$, 
where $\hat K = \hat K_{\rm cl} + \hbar \hat K_{\rm q}$. 
Here the operator $\hat K_{\rm cl}$ includes 
only the first order derivatives and describes the corresponding 
classical limit, while the other operator $\hat K_{\rm q}$ 
includes higher order derivatives and is responsible for quantum effects. 
For the model given by Eq. (\ref{Hnonlinear}) we get
\begin{eqnarray}
\frac{\partial f}{\partial \tau}  &=& i (1 + \bar{\mu} + 2 \bar{\mu} |\alpha|^2) 
 		\left( \alpha^{\star} \frac{\partial}{\partial \alpha^{\star}} - 
                       \alpha \frac{\partial}{\partial \alpha}
		\right) f +  \nonumber \\
&&	 i \bar{\mu} \left( (\alpha^{\star})^2 \frac{\partial^2}
{\partial (\alpha^{\star})^2} -
			       \alpha^2 \frac{\partial^2}{\partial \alpha^2}
			\right) f .
\label{observable}
\end{eqnarray}
In particular, for $\hat f= \hat a$ the evolution of $f(\tau)$ corresponds
to the evolution of the condensate matter-wave field
$\alpha(\tau) \equiv \langle \alpha | \hat a(\tau) | \alpha \rangle$. In this
case Eq. (\ref{observable}) can be solved exactly \cite{gennabook} :
\begin{equation}
\alpha(\tau) = \alpha \, e^{-i(1 + \bar{\mu}) \tau} 
\exp[ |\alpha|^2 (e^{-2 i \bar{\mu} \tau} - 1)] .
\label{alfaexact}
\end{equation}
The quantum evolution of this expectation value departs from the corresponding classical 
evolution 
$\alpha_{\rm cl}(\tau) = \alpha e^{-i \omega_{\rm cl}\tau} $ as
$\alpha(\tau)=\alpha_{\rm cl}(\tau) e^{-\tau^2/2 \tau_E^2}[1+O(\bar\mu\tau)+O(|\alpha|^2\bar\mu^3\tau^3)]$, where
$\tau_E$ is the Ehrenfest time-scale given by
\begin{equation}
\tau_{E} = \frac{1}{ 2 \bar {\mu} |\alpha|} .
\label{ehrenfest}
\end{equation}
The amplitudes of quantum and classical observables coincide at times multiple of the 
quantum recurrence time-scale
\begin{equation}
\tau_{R} = \frac{\pi}{\bar{\mu}} .
\end{equation}
Note that the quasi-classical limit, which is considered in this paper, corresponds to the 
following condition: $\tau_E/\tau_R=\varepsilon^{1/2}/2\pi\ll 1$. So, in what follows we will
be interested only in the region of parameters where $\tau_E\ll\tau_R$.
Quantum recurrences of the matter wave field of a BEC in the {\it pure} quantum regime 
$\alpha \approx O(1)$ (or $\varepsilon\approx 1$) in each lattice site were observed in 
\cite{greiner} at $\tau_{R} \approx 100$ ($t_{R}=0.55 {\rm ms}$), larger than the 
corresponding Ehrenfest time-scale $\tau_{E} \approx 15$. At the same time, quantum 
dynamical effects in the quasi-classical region 
of parameters have still not been observed in BECs.

Expressing $\alpha=\sqrt{J/\hbar} e^{-i \theta}$, we can rewrite Eq. (\ref{observable})
as
\begin{equation}
\frac{\partial f}{\partial \tau} =
(1 + 2 \mu_{\rm cl}) \frac{\partial f}{\partial \theta} +
2 \varepsilon \mu_{\rm cl} \frac{\partial^2 f}{\partial J \partial \theta} .
\label{qsingularity}
\end{equation}
The quantum term appears as a singular perturbation of the classical equation
because the small parameter $\varepsilon$ multiplies the higher order
derivative. Note that the quantum effects for observables vanish in two cases: 
(i) $\varepsilon=0$, which corresponds to the classical limit, and (ii) $\mu_{\rm cl}=0$, 
which corresponds to the quantum linear oscillator. The fact that
for nonlinear quantum systems  the terms with high order derivatives in the evolution 
equations for the density matrix and for the Wigner function represent a singular 
perturbation to the classical limit (Liouville function) is well-known. However, 
in spite of a large number of papers on this subject, from this fact it is still
unclear what are the conditions for the quantum-classical correspondence for 
{\it observables}.  
The solution (\ref{alfaexact}) of Eq. (\ref{observable}) for the observable 
$\alpha(\tau)$ (and also for an arbitrary observable \cite{gennabook}) demonstrates
that quantum effects (second order derivatives in Eqs. 
(\ref{observable},\ref{qsingularity})) 
represent a 
singular perturbation to the classical equation for {\it observables}, which includes 
only the first order derivatives and can be solved by the method of classical 
characteristics \cite{genna1,gennabook}. This results in a secular behavior of quantum 
corrections in the solution for the observable $\alpha(\tau)$  (\ref{alfaexact}). So, 
the question is: Under what conditions does the  environment ``kill'' (if at all) the 
quantum corrections which represent a singular perturbation to the 
{\it observables} of the classical world?    

Following \cite{milburn2,milburn3} we model the environment as a bath of harmonic 
oscillators in thermal equilibrium at a rescaled temperature
$\bar{\beta}=\hbar\omega/k_{\rm B} T$ linearly coupled through position 
to the QNO \cite{future}. In the Born-Markov approximation, the master equation for 
the reduced density matrix reads
\begin{equation}
\frac{d \hat \rho}{d \tau} = F_{\rm free}(\hat \rho) + 
F_{\eta}(\hat \rho) + F_{\nu}(\hat \rho) ,
\label{bm}
\end{equation}
where the first term, 
\begin{equation}
F_{\rm free}(\hat \rho) = -i [\hat a^{\dagger} \hat a + 
\bar{\mu} (\hat a^{\dagger} \hat a)^2, \hat \rho ] ,
\end{equation}
corresponds to the free, unitary evolution, the second one
\begin{eqnarray}
F_{\eta}(\hat \rho) &=& \frac{i}{2}  [\hat a+ \hat a^{\dagger}, 
\{ \hat{A}_1(\tau) \hat a + \hat a^{\dagger} \hat{A}_1(\tau) + \nonumber \\
&& i ( \hat{A}_2(\tau) \hat a - \hat a^{\dagger} \hat{A}_2(\tau)), \hat \rho \} ] ,
\end{eqnarray} 
accounts for dissipation, and the third one,
\begin{eqnarray}
F_{\nu}(\hat \rho) &=& - \frac{1}{2}  [\hat a+ \hat a^{\dagger}, 
[ \hat{B}_1(\tau) \hat a + \hat a^{\dagger} \hat{B}_1(\tau) + \nonumber \\
&& i ( \hat{B}_2(\tau) \hat a - \hat a^{\dagger} \hat{B}_2(\tau)), \hat \rho ] ] ,
\end{eqnarray} 
is related to noise. The time-dependent, operator-valued coefficients 
$\hat{A}_i$ and $\hat{B}_i$ depend on the frequency operator
$$
\hat{\Omega}=1 + \bar{\mu} (1+ 2 \hat{a}^{\dagger} \hat{a}),
$$ 
and on the spectral density of the environment  
$$J(\bar{\omega})=\frac{\gamma \bar{\omega} \bar{\Lambda}^2}
{\bar{\Lambda}^2 + \bar{\omega}^2},
$$
that we chose to be Ohmic,
 with
$\bar{\Lambda}$ a UV cut-off and $\gamma$ 
a system-environment coupling constant. Explicitly
\begin{eqnarray}
\hat A_1(\tau)&=& \int_0^{\tau} ds \eta(s) \cos(\hat \Omega s) ;\\
\hat A_2(\tau)&=& \int_0^{\tau} ds \eta(s) \sin(\hat \Omega s) ;\\
\hat B_1(\tau)&=& \int_0^{\tau} ds \nu(s) \cos(\hat \Omega s) ;\\
\hat B_2(\tau)&=& \int_0^{\tau} ds \nu(s) \sin(\hat \Omega s) ,
\end{eqnarray}
where the dissipation and noise kernels are respectively given by
\begin{eqnarray}
\eta(s)&=& \int_0^{\infty} d{\bar \omega} \frac{\bar \omega}{\pi} J(\bar \omega) 
\sin(\bar \omega s) ; \\
\nu(s) &=& \int_0^{\infty} d{\bar \omega} \frac{\bar \omega}{\pi} J(\bar \omega)
\coth(\frac{ \bar{\beta} \bar \omega}{2}) \cos(\bar \omega s).
\end{eqnarray}
The matrix elements of the operators $\hat A_i$ and $\hat B_i$ can
be straightforwardly computed in the Fock basis and shown to have an analogous
behavior to that of the $\bar{\mu}=0$, quantum Brownian motion case \cite{paz}.

\begin{figure}[t]
\includegraphics[scale=0.36]{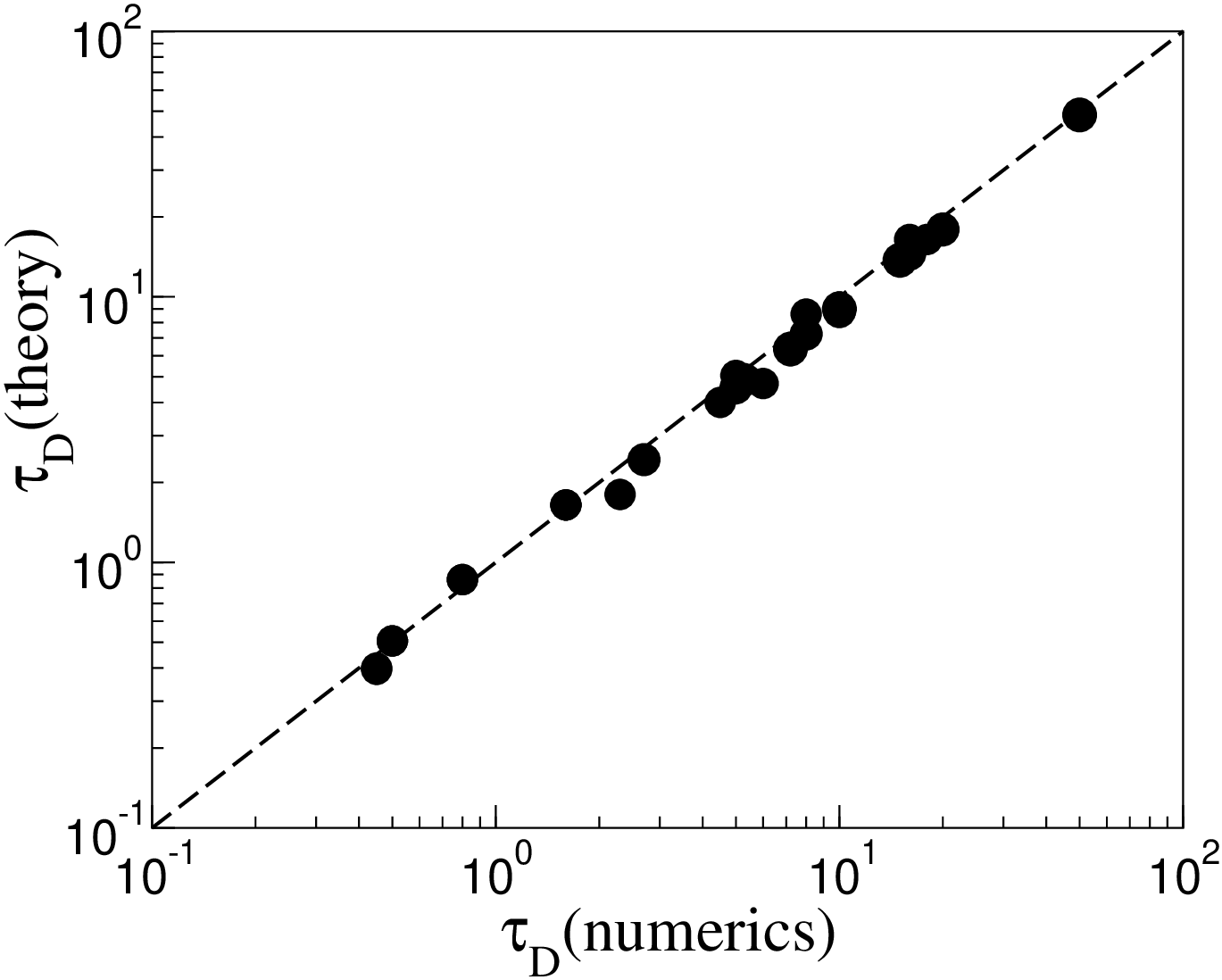}
\caption{
Comparison between the numerically obtained decoherence time $\tau_D$
and the approximate relation given by Eq. (\ref{taudeco}), obtained retaining only one
term in the master equation. The dashed line  corresponds to
$\tau_D({\rm theory})=\tau_D({\rm numerics})$.
Data are obtained by varying parameters in the following regions:
$20 \leq I_0 \leq 100$, $10^{-3} \leq \bar{\mu} \leq 4$, 
$10^{-2} \leq \bar{\beta} \leq 1 $, $10^{-5} \leq  \gamma \leq 10^{-2}$. 
}
\label{mil}
\end{figure}

To study the decoherence effects of the environment we start by considering 
an initial Schr\"odinger cat state formed by large-amplitude coherent states, 
$\hat \rho(0)= \cal{N} (|\alpha\rangle+|\beta\rangle) 
(\langle \alpha| + \langle \beta|)$, with  $|\alpha|^2, |\beta|^2 \gg 1$, 
and $\cal{N}$ a normalization constant. For simplicity we take $\alpha$
and $\beta$ to lie along a common radius, and we parameterize them as 
$\alpha=x e^{i \theta}$ and $\beta=(x+\delta x) e^{i \theta}$. 
Decoherence is due to the term in the master equation containing $\hat B_1$.
In the coherent state basis, the off-diagonal matrix elements of the density 
matrix evolve as
\begin{equation}
\frac{d}{d \tau} \langle \alpha | \hat \rho|\beta \rangle \approx - 
2 B_1(I_0,\tau) (\delta x)^2
\langle \alpha | \hat \rho | \beta \rangle ,
\label{rhom}
\end{equation}
where $B_1(I_0,\tau)=\langle \alpha | {\hat B}_1(\tau) | \alpha \rangle$ with
$I_0 \equiv |\alpha^2|$.
For $\tau \gg 1/\bar{\Lambda}$, $B_1(I_0,\tau)$ is approximately equal to
its asymptotic value 
$$
B_1(I_0,\infty) = \frac{\gamma \bar \Omega}{2}
\frac{ \bar \Lambda^2} {\bar \Lambda^2 + \bar \Omega^2}
 \coth\left(\frac{\bar \beta \bar \Omega}{2}\right),
$$
with $\bar \Omega=1+ \bar \mu(1+2 I_0)$. Therefore, the decoherence time-scale
is 
\begin{equation}
\tau_D= \frac{1}{2 B_1^n(\infty) (\delta x)^2}.
\label{ttd}
\end{equation}
This decoherence time-scale coincides
with the time-scale for exponential decay of quantum recurrences for an initial
coherent state $|\alpha \rangle$ coupled to the thermal bath. 
Indeed, in this case 
$(\delta x)^2 \approx \langle \alpha | x^2 | \alpha \rangle \approx I_0$, 
so that
\begin{equation}
\tau_D = \frac{\tanh(\bar{\beta} \bar{\Omega}/2)}{I_0 \gamma \bar{\Omega}} .
\label{taudeco}
\end{equation}
We checked this estimation by numerical simulations
of Eq. (\ref{bm}) using as initial state a quasi-classical coherent state.
As can be seen in Fig. \ref{mil}, the agreement is fairly good within the numerical errors.
In the limit of small temperature $\bar{\beta} \bar{\Omega} \gg 1$ 
and small non-linearity $\bar{\mu} I_0 \ll 1$ the decoherence time-scale 
coincides with the one derived in \cite{milburn2}; however, we 
stress that in the general case, differently from \cite{milburn2},
the decoherence time, as given by (\ref{taudeco}), depends on 
the parameter of nonlinearity $\bar{\mu}$.
In the high-temperature limit, $\bar{\beta} \bar{\Omega} \ll 1$, the decoherence 
time is: $\tau_D\approx\tau_\gamma \hbar^2\omega/4k_BTJ_{\rm cl}$, where
$\tau_{\gamma} = 2/ \gamma$ is the time scale of the relaxation of quantum observables
due to the interaction with the environment. 

\begin{figure}[t]
\includegraphics[scale=0.34]{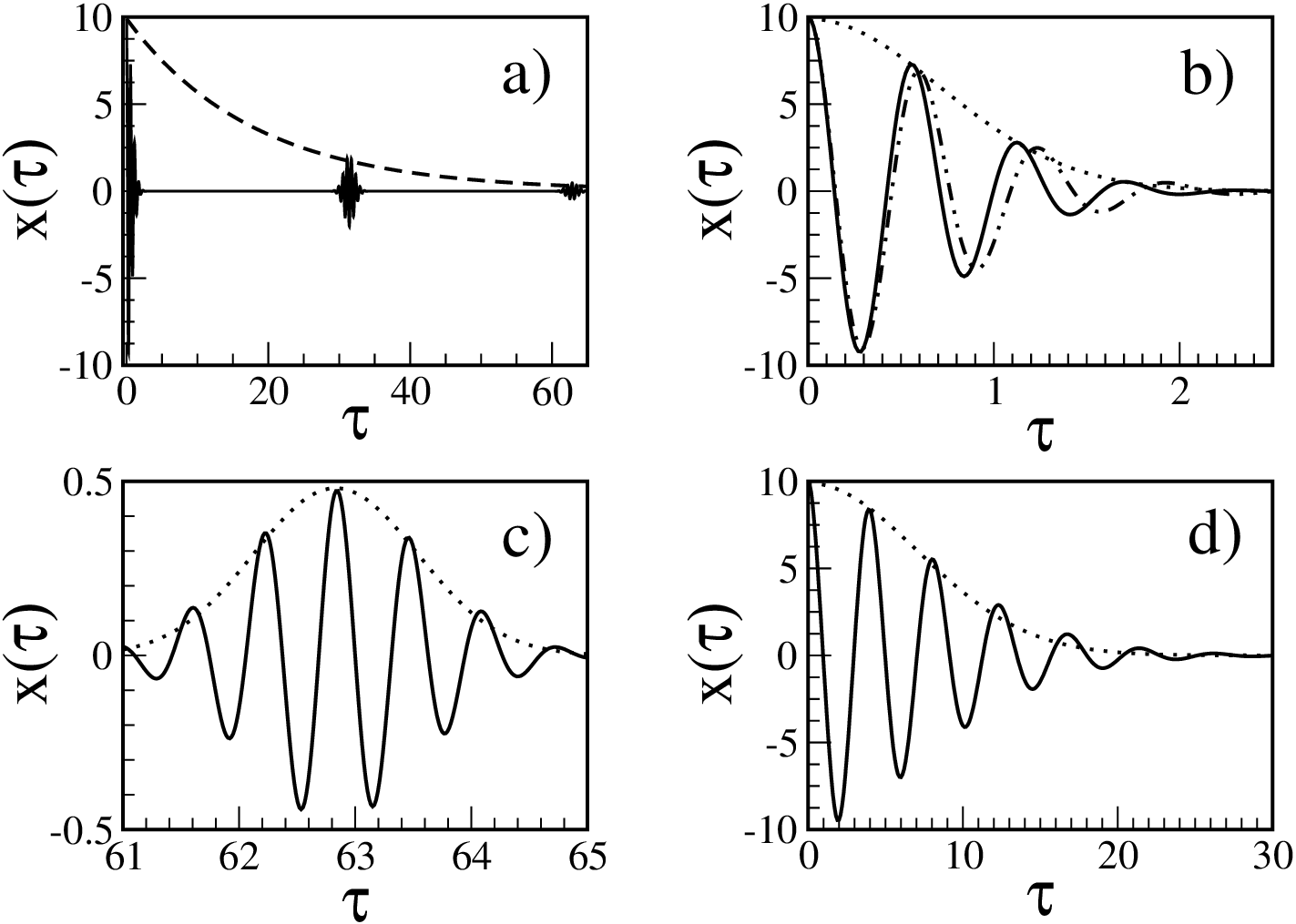}
\caption{
The average position as a function of the dimensionless time $\tau$.
In all cases $I_0=50$, $\bar{\beta}=1$. a) Parameters are $\bar{\mu}=0.1$ and
$\gamma=10^{-4}$. The dashed curve corresponds to $\exp(-\tau/\tau_D)$, where
$\tau_D=18> \tau_E \approx 0.7$. b) is an enlargement of the first bump of  a). 
Two more curves have been added: a dashed-dotted line corresponds to the average 
position for $\gamma=10^{-2}$, so that $\tau_D=0.18 < \tau_E$;
and a dotted line that corresponds to the envelope $\exp(-\tau^2/2\tau_E^2)$.
c) is an enlargement of the third bump of a). The dotted line corresponds to
the envelope $\exp(-\tau_R/\tau_D) \exp[-(\tau-\tau_R)^2/2\tau_E^2]$, where $\tau_R=10\pi$.
d) Parameters are $\bar{\mu}=\gamma=10^{-2}$, so that $\tau_D \ll \tau_E \ll \tau_{\gamma} < \tau_R$.
}
\label{dy}
\end{figure}

Having the density matrix elements, 
we can easily determine the average values of any observable, 
for instance, for the position
$x(\tau)\equiv \langle \hat x(\tau) \rangle = (\alpha(\tau) + \alpha^*(\tau))/\sqrt{2}$. 
 An example of such simulations is given in 
Fig. \ref{dy}. The time-scale for the overall decay of the amplitude of 
recurrences, shown in Fig. \ref{dy}a, is set by the decoherence time-scale 
$\tau_D$: the relative heights of two peaks, taken at two neighbor recurrent
times, is reduced by a factor $\exp(-\tau_R/\tau_D)$. 
An enlargement of  the first bump  of Fig. \ref{dy}a is given in Fig. \ref{dy}b. 
As one can see in both cases reported in the figure, 
the time-scale which governs the envelope of $x(\tau)$ is the Ehrenfest time 
$\tau_E$, which is independent of the coupling to the environment.
The same is true for any of the following revival bumps, see 
Fig. \ref{dy}c. Let us also notice that in Fig. \ref{dy}b the curves for $\gamma=10^{-4}$
(solid line) and for $\gamma=10^{-2}$ (dashed-dotted line) are slightly 
shifted one to the other.
This is due to the fact that the frequency of motion is slightly 
dependent on the bath-oscillator interaction strength,
$\omega_{\rm eff}^2 = \bar{\Omega}^2 - \gamma \bar{\Lambda}^3 / (\bar{\Lambda}^2+\bar{\Omega}^2)$.
This is not at all
surprising since the renormalization of the frequency is a feature of the 
considered master equation \cite{paz}. 

In Fig. 2d the average position is plotted for a case in which 
$\tau_D \ll \tau_E \ll \tau_{\gamma} < \tau_R$
($\tau_D=0.74$, $\tau_{\rm cl}=3.12$, $\tau_E=7.1$, $\tau_{\gamma}=200$ and $\tau_R=314$).
This figure represents the most important result of the present work: Usually,
in the quasi-classical region of parameters, and for rather large values of the coupling
to the environment ($\gamma \ge 10^{-2}$), the characteristic decoherence time-scale
is much shorter than the Ehrenfest time scale, $\tau_D \ll \tau_E$. 
Despite this, the system does not become entirely ``classical'', since quantum effects 
persist up to the Ehrenfest time. Indeed, for $x(0)=(\alpha+\alpha^*)/\sqrt{2}= \sqrt{2 I_0}=10$ 
and $\tau_D=0.74$, the dependence $x(0)\exp(-\tau/\tau_D)$ would give us, for example, for $\tau=10$ 
the value $1.3 \times 10^{-5}$, which is significantly  smaller than the 
correspondent value 3.7 defined by the function $x(0)\exp(-\tau^2/2\tau^2_E)$ for 
$x(0)=10$, $\tau=10$ and $\tau_E=7.1$ (which corresponds to the results presented in Fig. 2d.) 

\begin{figure}[t ]
\includegraphics[scale=0.32]{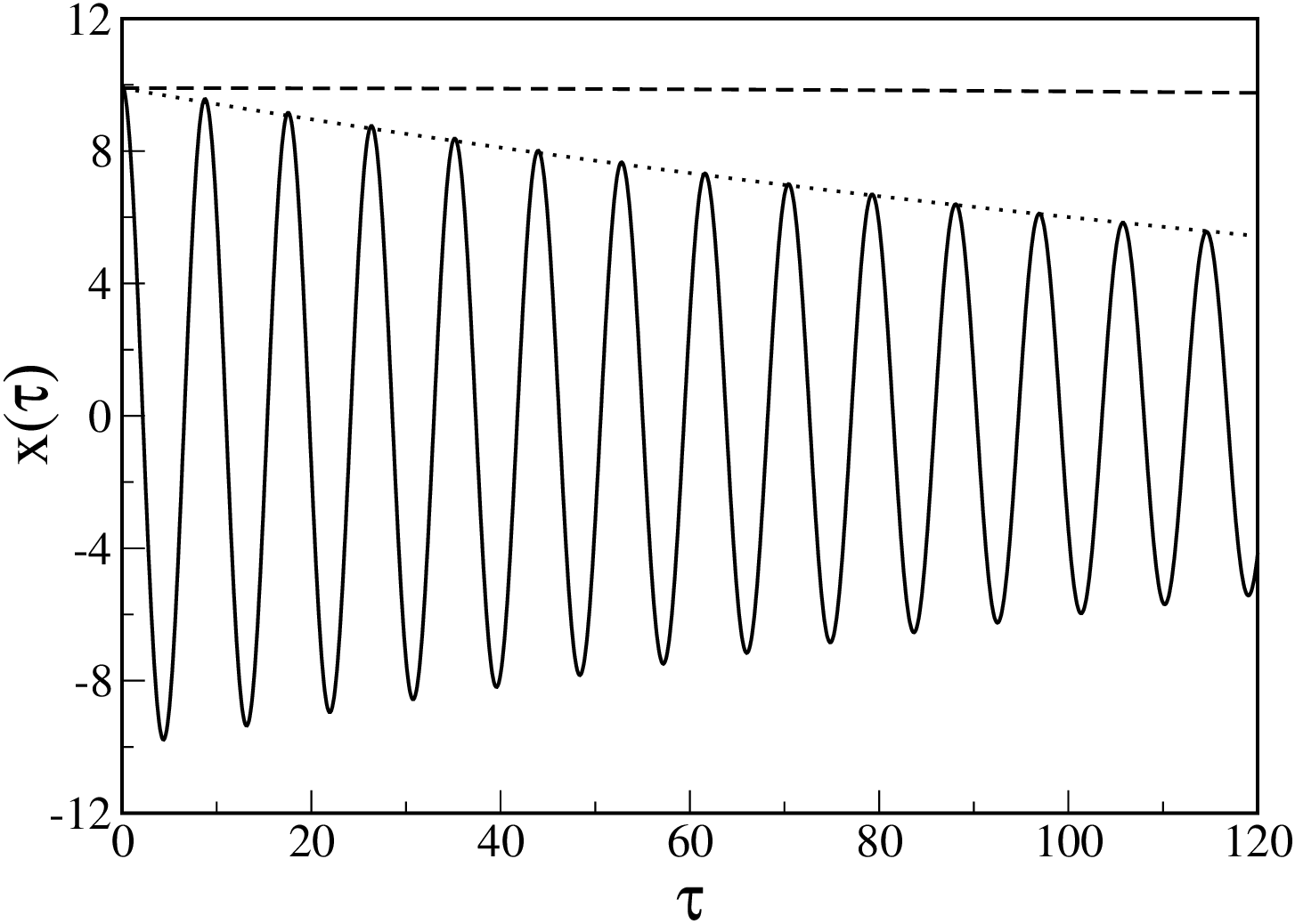}
\caption{
The average position $x(\tau)$
in the ``classical'' limit:  $\tau_D \ll\tau_{\rm cl} \ll \tau_\gamma \ll \tau_E\ll\tau_R$.
Parameters are: $\bar{\mu}=10^{-4}$, $\bar{\beta}=1$, $\gamma=0.01$, $I_0=50$,
so that $\tau_D=0.92$, $\tau_{\rm cl}=2\pi$, $\tau_\gamma = 200$, $\tau_E=707$, 
$\tau_R=\pi\times 10^4$.
The correct decay (dotted line), as given by
the relaxation time $\tau_\gamma$, is compared with the
``wrong'' decay (dashed line) given by the Ehrenfest time $\tau_E$.
}
\label{ss}
\end{figure}

In our model the classical limit corresponds to the following inequalities:
$\tau_D\ll\tau_{\rm cl}\ll\tau_{\gamma} \ll \tau_E \ll \tau_R$ (see Fig. \ref{ss}).
Because in the quasi-classical region of parameters the inequalities $\tau_E \ll \tau_R$
and $\tau_D \ll \tau_{\rm cl}$ are always satisfied, the really important condition for
the classical limit is $\tau_{\gamma} \ll \tau_E$.
In this case the system effectively behaves as a classical damped oscillator, and
quantum effects cannot be observed. For comparison
we plot in Fig.  \ref{ss} the overall decay of oscillation, given by $\tau_{\gamma}$,
with the one that would be given by the Ehrenfest time. The perfect agreement
of the decay relaxation time with the data and their wrong dependence on the Ehrenfest 
time is a manifestation of the classicality for this case.

To gain a qualitative understanding of these numerical findings we consider
a simplified version of our master equation Eq. (\ref{bm}) at zero
temperature, in which
we keep only the effect of dissipation and decoherence due to the environment,
and make the rotating wave approximation. In this way we get the standard
master equation in quantum optics for the QNO:
\begin{equation}
\frac{d\hat \rho}{d \tau} = -i [\hat a^{\dagger} \hat a + \bar{\mu}
(\hat a^{\dagger} \hat a)^2,\hat \rho] + 
\frac{\gamma}{2} (2 \hat a \hat \rho \hat a^{\dagger} -
\hat a^{\dagger} \hat a \hat \rho - \hat \rho \hat a^{\dagger} \hat a) .
\end{equation}
This equation is precisely the one considered in \cite{milburn2}, where an exact
solution for the quasi-probability $Q$ function was obtained, assuming an 
initial coherent state $|\alpha\rangle$. 
In particular, an exact expression for time-evolution of the average position 
$\langle \hat x(\tau)\rangle=(\langle \hat a(\tau) \rangle + 
\langle \hat a^{\dagger} (\tau) \rangle)/\sqrt{2}$ can be written, where 
\begin{equation}
\langle \hat a(\tau) \rangle = \alpha e^{(1+\bar{\mu})\tau}
e^{-\frac{\gamma \tau}{2}} e^{-\frac{|\alpha|^2}{1+k^2} (1+i k) 
(1-e^{- \gamma \tau} e^{-2 i \bar \mu \tau})} ,
\label{milburnsolution}
\end{equation}
with $k=\gamma/2 \bar{\mu}$. Note that for no coupling to the environment 
($\gamma=0$) we recover Eq. (\ref{alfaexact}). From Eq. (\ref{milburnsolution})
we can read the decay factor of the average position $x(\tau) \propto e^{-D(\tau)}$. It is
given by 
\begin{eqnarray}
D(\tau)&=& \frac{\gamma \tau}{2} + 
\frac{4 \bar\mu^2 |\alpha|^2}{4 \bar\mu^2 + \gamma^2} \times \\
&& \left[ (1-e^{-\gamma \tau} \cos(2 \bar\mu \tau)) - \frac{\gamma}{2 \bar\mu} 
e^{-\gamma \tau} \sin(2 \bar\mu \tau) \right] \nonumber .
\end{eqnarray}
Let us first analyze the case $\gamma/2 \ll \bar\mu$, corresponding
to $\tau_E \ll \tau_{\gamma}$.
Let us express the time $\tau$ around a given recurrence time as 
$\tau=n \tau_R + \tilde \tau$, where $n$ is a non-negative integer.
Assuming that the time $\tau$ is much shorter than the
relaxation time, $\gamma \tau \ll 1$, and that $\bar \mu \tilde{\tau} \ll 1$,
we can expand 
$D(\tau) \approx |\alpha|^2 ( 2 \bar\mu^2 \tilde{\tau}^2 + n \tau_R \gamma)$. 
Hence, in these limits, the decay of the average position is
\begin{equation}
x(\tau) \propto e^{-\tilde \tau^2/2 \tau_E^2} \; 
e^{-  n \tau_R/\tau_D},
\end{equation}
where $\tau_E$ is defined in Eq. (\ref{ehrenfest}) and 
$\tau_D=1/\gamma |\alpha|^2$ is the zero temperature limit of
Eq. (\ref{taudeco}) for $\bar \mu |\alpha|^2 \ll 1$. Therefore,
the decay of $x(\tau)$ within any recurrence bump is determined
by the Ehrenfest time-scale, and even in the limit $\tau_D \ll \tau_E$
the decay within the first bump ($n=0$) is still governed by the
Ehrenfest time-scale, which agrees with our numerical results presented above.
This implies that some quantum effects survive the loss of quantum coherence
due to the interaction with the environment. One the other hand, when $\gamma/2 \gg \bar \mu$ (i.e.,
$\tau_{\gamma} \ll \tau_E$) the classical limit is attained, and the decay
is governed by the relaxation rate. 

\begin{figure}[t]
\includegraphics[scale=0.36]{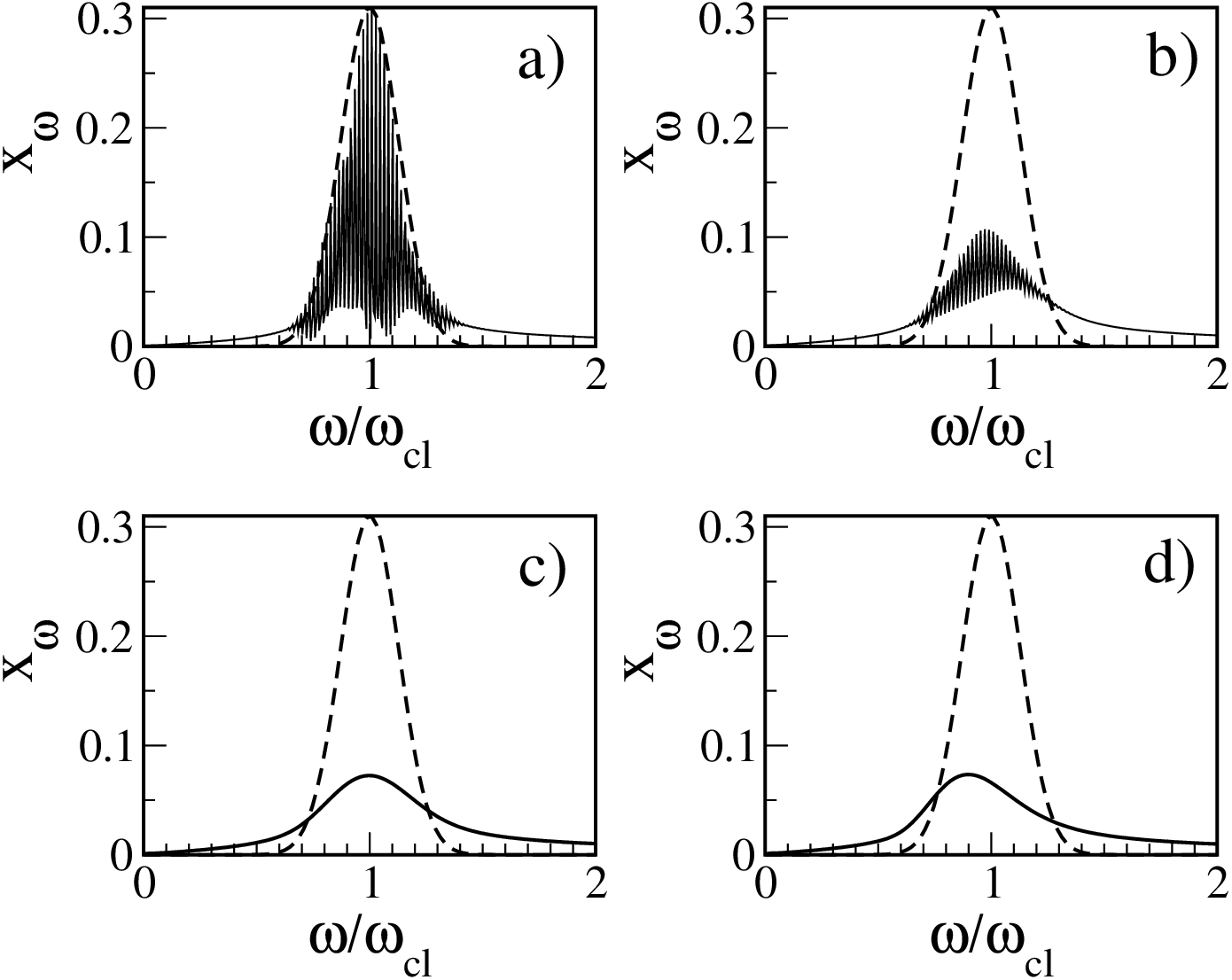}
\caption{
The Fourier spectrum of the average position versus the rescaled
frequency, where $\omega_{\rm cl}=1+2 \bar \mu I_0$ is the classical
frequency. 
As the dashed black line we show the theoretical expression
Eq. (\ref{erex}) for the closed system. 
Parameters $\bar{\beta}, \bar{\mu}, I_0$ are the same as in Fig. (\ref{dy}),
while 
a) $\gamma=10^{-5}$, $\tau_D=180 \gg \tau_E=0.7$,
b) $\gamma=10^{-4}$, $\tau_D=18 > \tau_E=0.7$
c) $\gamma=10^{-3}$, $\tau_D=1.8 > \tau_E=0.7$
d) $\gamma=10^{-2}$, $\tau_D=0.18 < \tau_E=0.7$
As one can see, decreasing of the decoherence time has
no effects on the width of the Fourier spectrum.
}
\label{fofo}
\end{figure}

Persistence of quantum effects after the decoherence time can be also
observed analyzing the Fourier spectrum of the average position in time
$x(\tau)=\sum_{\omega} x_{\omega} e^{i \omega\tau}$. When the system is closed, i.e.
no coupling to the environment, the Fourier components are given by
$$
x_{\omega}= \frac{\alpha}{2}  \Lambda\left( \frac{\omega-1}
{2 \bar{\mu}} \right) + {\rm c.c.},
$$
 where
\begin{equation}
\Lambda(n)={{\bar\mu}\over{\pi}} \int^{\pi/\bar\mu}_0 
e^{(e^{2i\bar\mu\tau}-1) |\alpha|^2 - 2i\bar\mu n\tau}d\tau .
\end{equation}
Estimating this quantity in the limit $\bar{\mu} \tau \ll 1$ 
we obtain the frequency spectrum for the QNO,
\begin{equation}
x_{\omega} \approx \frac{1}{\sqrt{4 \pi |\alpha|^2}} 
\exp \left[- \frac{(\omega-\omega_{\rm cl})^2}{2 \Delta\omega^2} \right] \ x(0) ,
\label{erex}
\end{equation}
which is a Gaussian distribution centered around the classical oscillation frequency 
$\omega_{{\rm cl}}=1+2\bar\mu|\alpha|^2$ with a spectral width given by the 
inverse of the Ehrenfest time-scale, $\Delta \omega = \tau_{E}^{-1}$.
In Fig. \ref{fofo} the Fourier spectrum $x_{\omega}$ for the QNO coupled
to the environment is shown for the case $\tau_E \ll \tau_{\gamma}$. 
As one can see, in all cases presented in Fig. \ref{fofo}
(whatever the relation is between the decoherence time and the Ehrenfest time)
the width of the Fourier spectrum is always given by the inverse Ehrenfest time.
When $\tau_{\gamma} \ll \tau_E$ the width of the spectrum is given by the
relaxation rate, and the classical limit is obtained.

The important condition for survival of the quantum effects for observables related to the 
Ehrenfest time-scale is $\tau_E \ll \tau_{\gamma}$, which can be written in the form
\begin{equation}
\Theta \equiv \frac{\tau_{\gamma}}{\tau_E} = 2 \mu_{\rm cl} \varepsilon^{1/2} 
\tau_{\gamma} \gg 1 .
\label{quantum}
\end{equation}
For BECs the parameter of nonlinearity $\mu_{\rm cl}=\mu J / \omega$ can be written as
$\mu_{\rm cl}= N \sqrt{a^2 m \omega / 2 \pi \hbar}$, where $N$ is the
number of particles in the condensate, $a$ is the s-wave scattering length, $m$
is the mass of the atoms and $\omega$ is the trapping frequency. The quasi-classical
parameter is $\varepsilon=1/N$. Therefore
\begin{equation}
\Theta_{\rm BEC} = a \sqrt{\frac{2 m \omega N}{\pi \hbar}} \tau_{\gamma} \gg 1 .
\end{equation}
For example, for $a=5 {\rm nm}$, $m=1.5 \times 10^{-25} {\rm kg}$, $\omega/2 \pi =100 {\rm Hz}$
and estimating the dimensionless relaxation time $\tau_{\gamma}$ from
the lifetime of the condensate (say $t_{\gamma}= 1 {\rm sec}$, so $\tau_{\gamma}=\omega t_{\gamma} = 2 \pi \times 10^2$), 
we need a total number of particles $N \gg 1$. 
In the case of a cantilever (or a mechanical resonator) 
the quasi-classical parameter is $\varepsilon=1/n$, where $n$ in the average number of levels involved in
the coherent state of the cantilever.
For the dimensionless relaxation time $\tau_{\gamma}$ we take  
$\tau_{\gamma}=2 Q$, where $Q$ is the cantilever quality factor. 
Then, for a cantilever 
the condition (\ref{quantum}) takes the form 
\begin{equation}
\Theta_{\rm cantilever}=\frac{4 \mu_{\rm cl} Q}{\sqrt{n}} \gg 1. 
\label{cantilever}
\end{equation}
We take the following dimensional parameters \cite{rugar1}: 
the amplitude of the cantilever oscillations $x_m=10$nm, the spring constant $
k_c=6\times 10^{-4}$N/m and the frequency of the fundamental mode of the cantilever 
$\omega_c/2\pi=6.6$kHz. In this case, the number of cantilever levels can be estimated 
as $n\approx k_cx^2_m/\hbar\omega_c\approx 6\times 10^{11}$. We also take $Q=10^6$. 
Then, we have from 
Eq. (\ref{cantilever}) the estimate for the parameter of nolinearity: $\mu_{\rm cl}\gg 0.2$.
 We hope that these conditions can be experimentally realized, and quantum effects related to the 
Ehrenfest time-scale can be observed in the quasi-classical region of parameters.

We acknowledge useful discussions with G. Nardelli. This work was supported by the Department of Energy (DOE) under
Contract No. W-7405-ENG-36, and by the Defense Advanced Research Projects Agency (DARPA).

\end{document}